# High emissivity, thermally robust emitters for high power density thermophotovoltaics

Minok Park[1,2,3,†], Shomik Verma[4,†], Alina LaPotin[4], Dustin P. Nizamian[5], Ravi Prasher[1,2], Asegun Henry[4,*], Sean D. Lubner[1,6,*], Costas P. Grigoropoulos[1,2,*], and Vassilia Zorba[1,2,*]

[1]Energy Technologies Area, Lawrence Berkeley National Laboratory, Berkeley, California, 94720, USA
[2]Department of Mechanical Engineering, University of California at Berkeley, Berkeley, California, 94720, USA
[3]Department of Mechanical and Automotive Engineering, Kongju National University, Cheonan, 31080, Republic of Korea
[4]Department of Mechanical Engineering, Massachusetts Institute of Technology, Cambridge, Massachusetts, 02139, USA
[5]Antora Energy, Inc., Sunnyvale, California, 94089, USA
[6]Department of Mechanical Engineering, Division of Materials Science and Engineering, Boston University, Boston, Massachusetts, 02215, USA

[*]Corresponding authors: vzorba@lbl.gov; cgrigoro@berkeley.edu; slubner@bu.edu; ase@mit.edu
[†]These authors contributed equally to this work.

**Abstract**
Thermal radiative energy transport is essential for high-temperature energy harvesting technologies, including thermophotovoltaics (TPVs) and grid-scale thermal energy storage. However, the inherently low emissivity of conventional high-temperature materials constrains radiative energy transfer, thereby limiting both system performance and technoeconomic viability. Here, we demonstrate ultrafast femtosecond laser-material interactions to transform diverse materials into near-blackbody surfaces with broadband spectral emissivity above 0.96. This enhancement arises from hierarchically engineered light-trapping microstructures enriched with nanoscale features, effectively decoupling surface optical properties from bulk thermomechanical properties. These laser-blackened surfaces (LaBS) exhibit exceptional thermal stability, retaining high emissivity for over 100 hours at temperatures exceeding 1000°C, even in oxidizing environments. When applied as TPV thermal emitters, Ta LaBS double electrical power output from 2.19 to 4.10 W $cm^{-2}$ at 2200°C while sustaining TPV conversion efficiencies above 30%. This versatile, largely material-independent technique offers a scalable and economically viable pathway to enhance emissivity for advanced thermal energy applications.

## Introduction

High emissivity in the visible to infrared (IR) spectral range is crucial for effective thermal energy transport in solar and high-temperature energy applications, including thermophotovoltaics (TPVs)[1-3], concentrating solar power (CSP)[4,5], spacecraft thermal management[6,7], and solar water desalination[8-10]. In these systems, low emissivity materials can significantly limit a system's performance and economic viability, necessitating larger surface areas and increasing material and infrastructure costs[11-13]. Improving emissivity could thus have transformative effects on technologies that depend on thermal radiative energy transfer, especially in TPVs where enhanced power density and reduced system size could facilitate grid-scale thermal energy storage (TES) solutions that enable switching entirely to intermittent renewable energy[14].

A blackbody (BB) is defined as a perfect absorber and therefore emitter across all wavelengths and for all directions, regardless of light polarization[15-17]. Hence, a BB has unity emissivity, maximizing thermal radiative energy transfer for a fixed surface area. Achieving this standard for TPV applications would substantially increase power density and reduce cost per power, which is often the dominant cost in TPV-based heat recovery and TES systems[13].

Some conventional approaches for creating high-temperature broadband emitters use carbon-based materials with naturally high emissivities. However, these also have high vapor pressures that cause emitter sublimation and subsequent optically thick deposition onto the cooled TPV cells, rapidly degrading performance[18,19]. Another example, silicon carbide, can only be machined with diamond tipped tools[20]. Alternatively, lower vapor pressure non-carbon materials have been applied as coatings that achieve broadband absorption through geometric design, such as hierarchical structures, multilayer materials, or metamaterials[6,21-23]. However, these are prone to delamination and substrate compatibility challenges, such as from mismatched coefficients of thermal expansion or poor chemical adhesion, reducing practicality[24].

Specifically for TPV systems utilizing back-surface reflectors, no contemporary emitter simultaneously satisfies the following five metrics: optical performance (defined in this work as broadband high emissivity), scalability to large areas, high-temperature stability, system integrability, and cost[25]. Refractory materials could be ideal candidates for TPV emitters as they offer thermal durability[22], low vapor pressure[18], and good machinability[26], but they suffer from low intrinsic emissivities (e.g., Ta has an emissivity < 0.1 for wavelengths above 1 μm[27]). Decoupling these optical properties from the other metrics would significantly expand the viable design space, enabling the use of low vapor pressure and scalably machinable refractories. While this decoupling could in principle be accomplished via direct surface texturing, such approaches have thus far been limited to select materials and have not yet been demonstrated to produce significant and stable improvements to broadband emissivity at high temperatures[28-32].

Ultrafast femtosecond (fs) laser ablation, which can induce surface material removal through rapid phase transformation[16,33], has been widely used to alter surface structures for applications such as controlling wettability[34,35], drilling and patterning[36-38], structural surface coloration[39,40], and bio-materials[41,42]. While fs laser processing has previously been used to create micro/nanoparticle-decorated structured surfaces with metal oxide layers and plasmon hybridizations to enhance light absorption and emissivity[8,43], most efforts primarily focused on improving absorption within relatively narrow and specific spectral ranges (e.g., only up to the near-IR or solely in the mid-IR), while also frequently utilizing materials that are not suitable for high temperature applications due to their low melting points (e.g., ~660°C for aluminum)[44-48].

Here, we extend ultrafast fs laser ablation to more generally decouple surface broadband optical properties from thermomechanical stability across a variety of materials, creating persistent near-BB surfaces in high-temperature stable materials for TPV applications. A scalable fs laser ablation technique[49-51] is designed for fabrication of broadband high-emissivity, near-BB surfaces, termed laser-blackened surfaces (LaBS). We show that LaBS achieve near-unity spectral emissivity (> 0.96) across 0.3 to 15 μm on different substrates, including refractory materials, primarily through the formation of light-trapping microstructures, effectively decoupling optical properties from bulk characteristics. As a result, high emissivity (~0.93) is maintained after heating to temperatures of 1000°C in air and 1500°C in Argon. When used as a TPV thermal emitter from 1700°C to 2200°C, a Ta LaBS achieves a twofold increase in generated electrical power compared to untreated Ta, which could significantly improve TES scalability by reducing system size and cost per Watt of delivered stored energy.

Beyond TPVs, this fs laser processing method offers a largely material-independent approach to enhancing high-temperature emissivity for a range of applications where different bulk material properties are required. By decoupling surface emissivity from bulk material properties, this approach opens new possibilities for economically viable, high-temperature radiative surfaces in the energy sector.

**Ultrafast fs laser-material interactions for fabricating microstructures with nanoscale features**

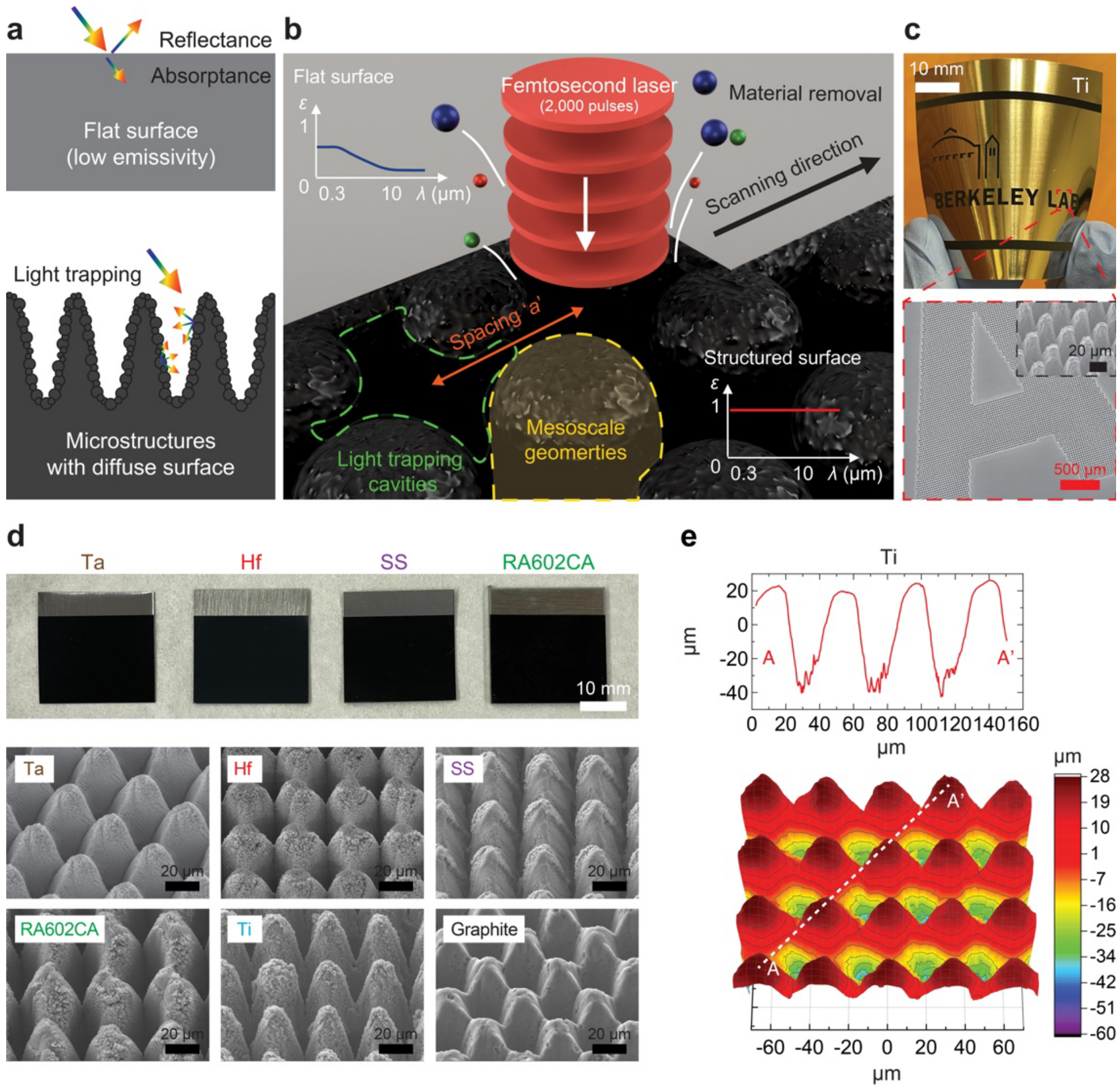

**Figure 1. Fabrication of mesoscale structures via fs laser processing.** (a) Light-material interaction with flat surfaces versus mesoscale geometries. Incident light is scattered and trapped by surface structures, producing additional light-material interactions that enhance total light absorption. (b) Schematic depicting the fabrication of near-BB surfaces using ultrafast fs laser fabrication. (c) Example of selective surface material texturing (dark regions) on a Ti foil (140 μm thickness). The SEM image shows the corresponding microstructure within the letter 'A'. (d) Different substrates with mesoscale surface structures, characterized by digital (top) and SEM (bottom) images demonstrating scalability and versatility of the process. The black areas in the digital images represent the laser-processed area, while the grey areas represent the pristine substrate. (e) Surface morphology of microstructures on Ti characterized by white light interferometry. The white scale bar is 10 mm, the black scale bar is 20 μm, and the red scale bar is 500 μm.

**Figure 1** describes the basis for enhanced light absorption to achieve near-BB surfaces. For a non-transparent optically flat surface at the interface with the air (**Fig. 1a, top**), the incident light is either reflected or absorbed as a function of complex refractive index (see details in Supplementary Note 1)[15,16]. Any deviation in the material's spectral refractive index from the ambient environment refractive index (1 for gas and vacuum) produces reflection and correspondingly yields emissivity lower than 1 for that wavelength. The spectral emissivity of metals in the IR regimes is typically low (i.e., they are of high reflectivity above 1 μm wavelength) because their dielectric functions increase with wavelength (and inversely decrease with respect to frequency). As a result, a flat metal surface cannot offer augmented thermal emission in the IR wavelength regime.

On the other hand, microstructured surfaces of multiscale texture and roughness can scatter and trap light within microcavities[7,52], leading to enhanced light absorption (**Fig. 1a, bottom**). To fabricate such surface morphologies, an ultrafast fs laser is employed for selective ablation of material from the surface, as depicted in **Fig. 1b-c** and Supplementary Fig. 1a. Specifically, 500-fs laser pulses at a wavelength of 1030 nm are focused on the target surface in ambient air. The incident laser power is fixed at 1.5 W with a repetition rate of 100 kHz and individual laser pulse fluences set to 2.1 J $cm^{-2}$. A total of 2,000 fs laser pulses are irradiated at a fixed location, and the same procedure is repeated at the next location separated by a spacing '*a*', to fabricate periodic surface structures on the entire surface area (Supplementary Fig. 1b and Supplementary Video 1–3). For applications operating in the range of 1000°C to 2000°C, the corresponding peak BB radiation wavelengths, $\lambda_{max}$, are respectively from 2.28 μm to 1.28 μm per Wien's displacement law. To choose the spacing, we considered the focused beam spot size of 30 μm, so the structural spacing remained below 35 μm to guarantee that the entire area was processed by the laser. Further, to ensure the fabricated structures are sufficiently larger than $\lambda_{max}$ to serve as efficient absorbers and emitters, the spacing '*a*' is chosen between 35 μm and 25 μm, depending on thermophysical properties of materials. Note that spacings smaller than 25 μm can be achievable under the same experimental conditions, depending on the target application and associated temperature ranges.

Upon irradiation of sufficiently high fluence pulses, the metal surface experiences complex phase transitions whose coupled dynamics result in material removal in the forms of a plasma plume, nanoparticles, and melt ejecta streaks[33,53,54]. Subsequently, remnant and redeposited melt matter is re-solidified into hierarchical microstructures featuring micro-/nano- particles, as characterized by scanning electron microscopy (SEM) images in **Fig. 1d** and Supplementary Fig. 2. Similar surface geometries can be obtained on different types of substrates using laser processing parameters in the same range as in the case of Ti. Due to the shallow absorption depth and the rapid energy deposition time of 500 fs, the extent of the heat affected zone into the irradiated material is limited. Furthermore, the temporal separation between successive pulses (i.e., 10 μs) is much longer than the time scale of surface melting and freezing, and therefore prevents heat accumulation in the irradiated target. Similarly, the ablation plume dynamics also evolve at a faster time scale (~μs)[53]. Consequently, the ablation process is in essence digital, eventually yielding microcavity recess depths in the range of 40 μm to 60 μm (aspect ratio ranges from 1.1 to 1.7), as measured by white light interferometry (WLI) in **Fig. 1e** and Supplementary Fig. 3. To achieve greater depths than 60 μm, increasing the laser fluence beyond the 2.1 J/$cm^2$ and applying more than 2000 laser pulses may be considered, depending on material.

This ultrafast fs laser fabrication technique is more versatile than application of surface coatings which are often bespoke to a specific material and face issues with high-temperature stability and adhesion[23,55]. The specific mesoscale surface geometries created with this technique are also difficult or impossible to create using other physical or chemical patterning methods such as reactive ion etching[56]. The fs laser processing technique is a simple, single-step, pigment and chemical-free method, and is rapid and scalable[49-51]. Specifically, the processing speed for the benchtop system used in this work is 13.9 min $cm^{-2}$, but further improvements as presented in Supplementary Note 2 and Supplementary Fig. 4a can accelerate processing tenfold to ~1 min $cm^{-2}$. Likewise, at benchtop scales (~10 $m^2$) the processing cost is approximately \$2 $cm^{-2}$, but at an industrial scale of 100,000 $m^2$ the projected processing cost drops considerably to 0.04 ¢ $cm^{-2}$ (Supplementary Fig. 4b).

**Broadband augmented spectral emissivity enabled by LaBS**

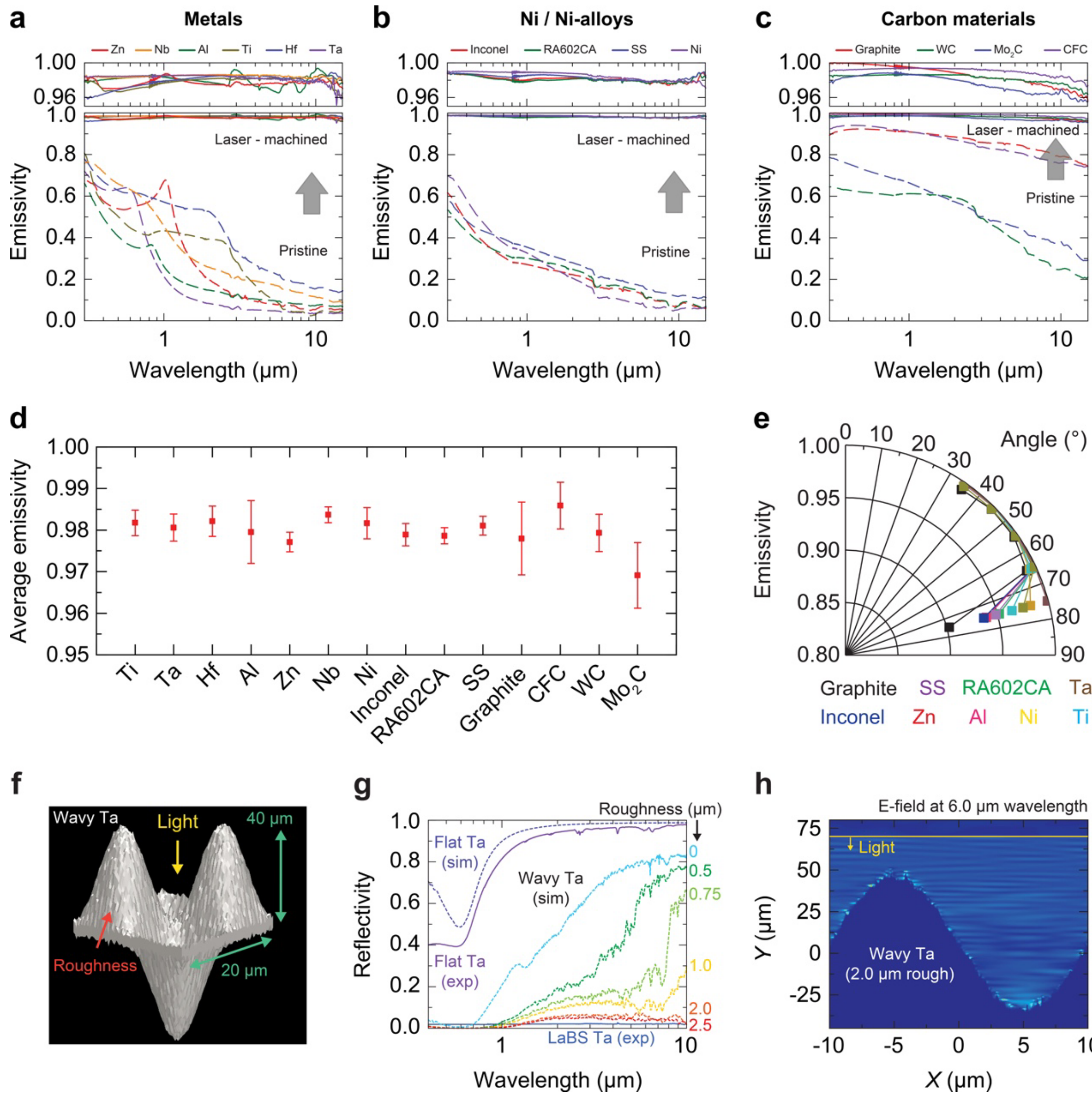

**Figure 2. Optical properties of LaBS.** Hemispherical spectral emissivity of LaBS (solid lines) and pristine flat surfaces (dashed lines) for (a) metals (Zn, Nb, Al, Ti, Hf, Ta), (b) Ni/Ni-alloys (Inconel, RA602CA, SS, Ni), and (c) carbon materials (graphite, CFC, WC, $Mo_2C$). Standard deviations across the 0.3 to 15 μm wavelength range are below 0.42% (0.0042 in emissivity), as shown in Supplementary Fig. 5. (d) Spectrally integrated average emissivity of LaBS within 0.3 μm to 15 μm wavelength. (e) Directional average emissivity of LaBS. (f) Layout of wavy Ta topography with surface roughness, for input to FDTD simulations. (g) Spectral reflectivity predicted by FDTD simulations (dotted lines; "sim") compared with experimental measurements (solid lines; "exp"). (h) Representative electric field at 6 μm wavelength for wavy Ta with 2 μm root-mean-squared surface roughness from FDTD simulations.

**Figure 2a-c** and Supplementary Fig. 5-6 shows the hemispherical spectral emissivity and its standard deviation across the 0.3 to 15 μm wavelength range for both pristine surfaces and laser-textured microstructures of various materials, including metals (Ta, Ti, Al, Hf, Zn, Nb), Ni/Ni-alloys (Ni, stainless steel, Inconel 600, RA602CA), and carbon containing materials (graphite, carbon fiber composites (CFC), tungsten carbide (WC), molybdenum carbide ($Mo_2C$)). Pristine metal/metal alloys exhibit a spectral emissivity lower than 0.8 at short wavelengths that monotonically decreases as the wavelength increases. Pristine carbon materials exhibit a nearly flat, gray emissivity until the mid-IR, followed by a gradual decrease throughout the extended IR range. Nevertheless, near-BB emissivity is observed on all the laser

textured materials due to the development of the hierarchical surface morphologies shown in **Fig. 1d** and Supplementary Fig. 2. These surfaces have spectral emissivity higher than 0.96 with standard deviations below 0.0042 over the same wavelength range after fs laser processing, with an average emissivity exceeding 0.97 (**Fig. 2d**). In addition, angle-resolved average emissivity measurements indicate that laser fabricated microstructures can absorb incident light up to 75-degree angle (**Fig. 2e**). Furthermore, we demonstrate that the spectral emissivity for a 200 μm thick Inconel LaBS is preserved even after folding at a 90-degree angle with a large radius of curvature (Supplementary Fig. 7). Given their optical behaviors closely resembling those of BBs[15,17], we refer to them as laser-blackened surfaces (LaBS).

To elucidate the origin of LaBS' optical properties, we conduct an analysis of their light-material interaction using Finite-Difference Time-Domain (FDTD) electromagnetic simulations using the software *Lumerical*. Wavy Ta microstructures (20 μm width and 40 μm height) with different surface roughnesses, mimicking laser induced cavities decorated with micro-/nano- particles, are investigated (**Fig. 2f**). For flat Ta with a smooth surface, both calculation and experimental measurement show a spectral reflectivity higher than 0.8 above 1 μm wavelength (**Fig. 2g**). For wavy Ta structures with a smooth surface, absorption below 0.7 μm wavelength (visible spectrum) is prominent, while IR light is still strongly reflected due to higher impedance mismatching[27] with air at longer wavelengths (Supplementary Fig. 8). The introduction of surface roughness enhances light absorption, and the wavy Ta substrate with 2 μm roughness suffices to present a near-BB surface up to 10 μm wavelength (average emissivity of 0.95), which aligns well with experimental results (0.98 for a surface roughness of 2.22 ± 0.4 μm shown in Supplementary Fig. 9). **Fig. 2h** and Supplementary Video 4 show the electric field near the surface and clearly indicate that the incident light is trapped, scattered, and absorbed within microcavities due to the presence of surface roughness, compared to smooth wavy Ta structures. Moreover, the light-material interaction for flat Ta with different surface roughness is studied (Supplementary Fig. 10). Under the same surface roughness of 2 μm, light of longer than 7 μm wavelength is not fully absorbed (spectral reflectivity of 0.2 at 10 μm wavelength), further highlighting the significance of hierarchical microstructures in achieving near BB surfaces.

Previous studies have indicated that the formation of oxide layers on surface geometries during fs laser processing can contribute to enhancing absorption in IR wavelengths, specifically for Al and steel substrates[43,44]. Our analysis using cross-sectional SEM and energy-dispersive X-ray spectroscopy confirms the presence of oxygen-containing layers of 2 to 4 μm thickness on Ti, Ta, and Hf microstructures (Supplementary Fig. 11). Oxide layers can contribute to increased absorption in certain IR ranges through phonon-polariton absorption[44,57,58] and help reduce refractive index mismatch. However, for our systems they are not the dominant cause of broadband near-BB emissivity. FDTD simulations show that the presence of oxide layers on microstructures has a measurable but smaller impact on the heightened light absorption compared with surface structuring effects (Supplementary Fig. 12). This is primarily attributed to the lower extinction coefficient of oxide materials in comparison to the high extinction coefficient of pure metals in the IR wavelength range[27] (0.24 for $Ta_2O_5$, and 54 for Ta at 10 μm wavelength). Therefore, combining the theoretical analysis in **Fig. 2g**, we conclude that the major contributor to LaBS on different types of materials is the textured light-trapping cavities, consisting of micro-/nano- particles as well as microstructures.

## Enhancement in thermal emission and TPV electrical power density

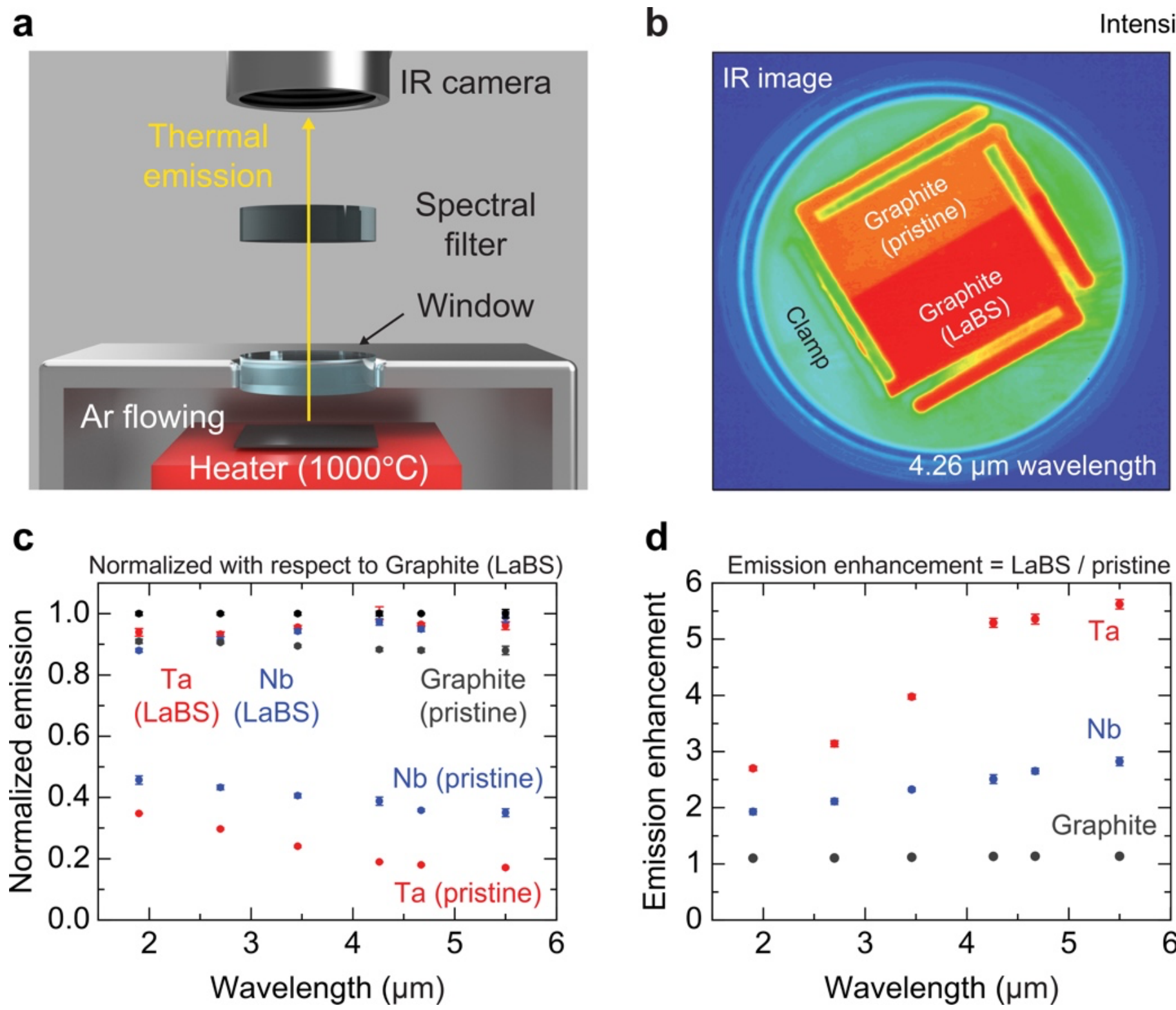

**Figure 3. Augmentation of thermal emission in LaBS compared to pristine surfaces.** (a) Schematic of thermal emission measurement using an IR camera and spectrally selective filters. The target samples are heated to 1000°C inside an argon flowing chamber using a resistive heater. Thermal emission is quantified using a photon counting method. (b) An example of acquired IR images for graphite at 4.26 μm wavelength. (c) Thermal emission for Ta and Nb LaBS, as well as pristine Ta, Nb, and graphite, normalized with respect to graphite LaBS. (d) Emissive power enhancement between LaBS and pristine surfaces.

TPVs are a promising emerging solid-state heat engine technology with broad applications including thermal energy storage[14,59], portable power generation[60,61], and waste heat recovery[62,63]. Because TPVs' primary mechanism of energy transport and conversion is based on high-temperature thermal radiation, they are an ideal technology to benefit from LaBS. Our specific emphasis is on enhancing the electrical power density of a TPV cell, governed by both the emitter's spectral emissivity and temperature.

It is important to first verify that an enhancement in room temperature spectral emissivity can result in increased thermal emission at elevated temperatures. We therefore quantify thermal emission enhancement for LaBS as compared to pristine surfaces at a temperature of 1000°C (**Figure 3a**). A set of IR camera and spectral filters is used to measure the thermal emission intensity using a photon counting method at specific wavelengths (refer to details in Supplementary Fig. 13). **Fig. 3b and c** shows the thermal emission normalized to graphite LaBS. Emission for Ta and Nb LaBS surpasses that of pristine graphite (gray points) across most of the measured spectral range and closely approaches graphite LaBS (black points) within the wavelength range of 1.9 to 5.5 μm. In contrast, pristine Ta and Nb surfaces emit less than half this amount. This observation indicates that laser textured metallic substrates can emit spectrally flat, 'gray' thermal radiation like the well-known behavior of graphite in the IR wavelength range. This characteristic is not usually accessible for pristine metal surfaces, as discussed in **Fig. 1**. The calculated emission enhancement factors (ratios between LaBS and pristine surfaces) shown in **Fig. 3d** quantify thermal emission increases of 170% to 460% for Ta, 100% to 170% for Nb, and 10% for graphite, depending on the considered wavelength.

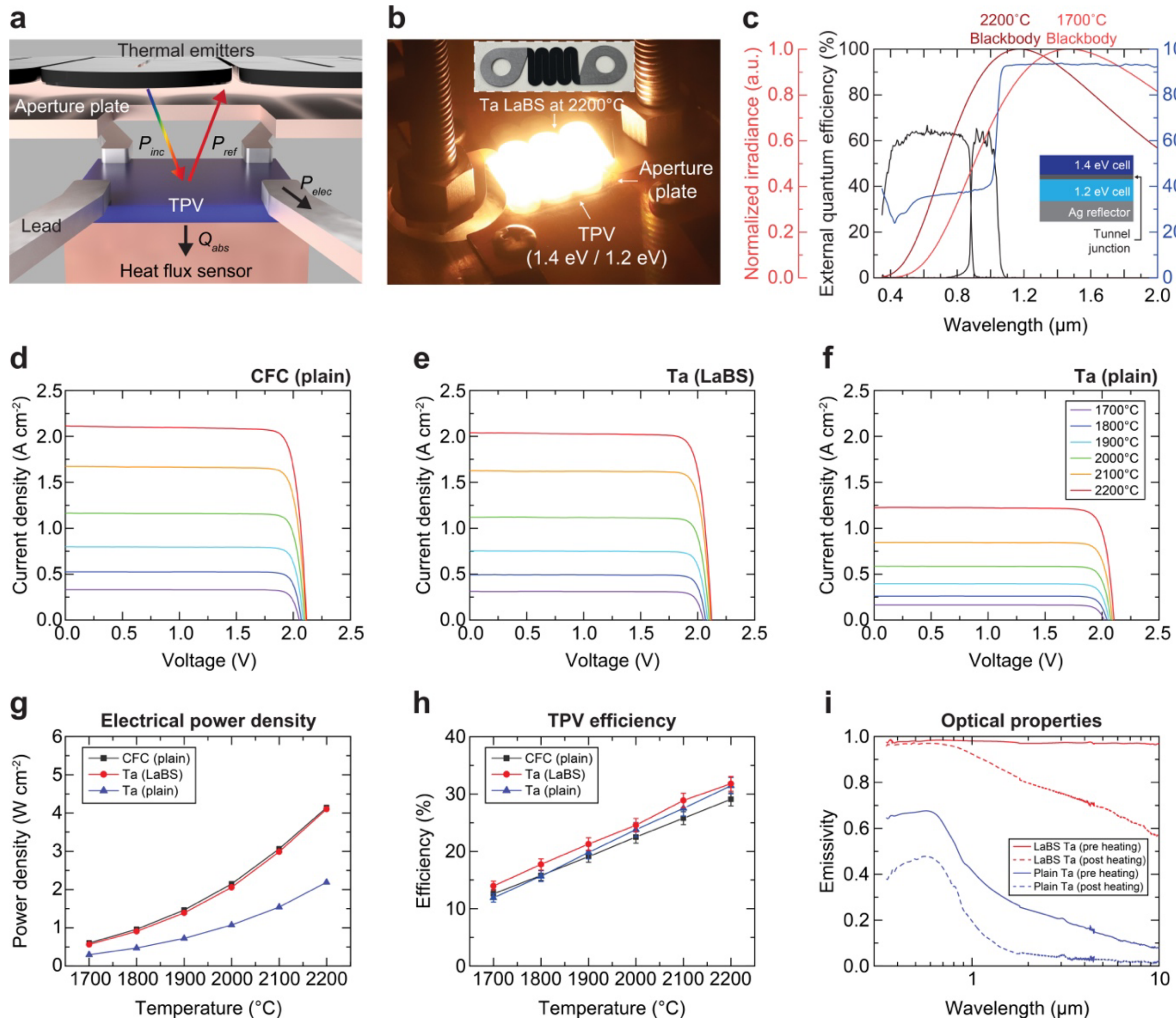

**Figure 4. Characterization of a tandem TPV cell with 1.4/1.2 eV bandgaps using different thermal emitters.** (a) Schematic of TPV characterization[64,65] with Ta LaBS. (b) Picture of the TPV characterization system with the Ta LaBS at 2200°C functioning as the TPV thermal emitter (inset: Ta LaBS at room temperature). (c) External quantum efficiency and reflectance of the 1.4/1.2 eV tandem TPV cell, with normalized BB emission spectra at 2200 °C and 1700°C overlaid. The inset illustrates a simplified schematic of the TPV cell structure with the full schematic available in LaPotin et al.[1]. Current density-voltage (IV) curves for TPV cells powered by (d) CFC, (e) Ta LaBS, and (f) plain Ta, respectively, for emitter temperatures ranging from 1700°C to 2200°C. (g) Generated electrical power density at maximum power point of the TPV cell as a function of emitter temperature. (h) TPV heat-to-electricity energy conversion efficiency measurements, error bars indicate 1 standard deviation uncertainties in heat flux measurements. (i) Optical property measurements for Ta LaBS and plain Ta before and after heating.

We experimentally demonstrate that the increase in thermal emission due to LaBS enhances the generated electrical power density in two tandem TPV cells with bandgaps of 1.4/1.2 eV (**Figure 4**) and 1.2/1.0 eV (Supplementary Fig. 14). We place thermal emitters directly above the TPV cells (**Fig. 4a-b**) and measure the electrical power density and heat absorbed as a function of emitter temperature. We test different thermal emitters including CFC, Ta LaBS (Supplementary Fig. 15 and Supplementary Video 5), and plain Ta in the temperature range of 1700°C (peak energy of 0.84 eV) to 2200°C (peak energy of 1.06 eV), selected to mimic TPV application-relevant temperatures. As indicated by the external quantum efficiency and reflectance curves in **Fig. 4c**, photons emitted above the lower bandgap (either 1.2 eV or 1.0 eV) can be absorbed by the TPV cell and converted to electricity, while unusable photons emitted below the bandgap are back-reflected and recycled at the emitter. Accordingly, the greater the number of photons emitted above the bandgap, the higher the electrical current produced by the same TPV.

**Fig. 4d-f** and Supplementary Fig. 14c-e show generated electrical current density versus voltage (IV) measurements of the TPV cells with a cell area of 0.7145 $cm^2$, using different emitters set at various temperatures. For the 1.4/1.2 eV TPV cell, the short circuit current density using the CFC emitter is 2.11 A $cm^{-2}$, Ta LaBS is 2.04 A $cm^{-2}$, and plain Ta is 1.22 A $cm^{-2}$ at 2200°C. Moreover, the power density calculated by dividing the electrical power at the maximum power point by the TPV cell area is shown in **Fig. 4g**, and power density versus voltage plots are presented in Supplementary Fig. 16-17. At 2200°C, the power density using the CFC emitter is 4.14 W $cm^{-2}$, Ta LaBS is 4.10 W $cm^{-2}$, and plain Ta is 2.19 W $cm^{-2}$. Similar enhancements are observed when using the 1.2/1.0 eV TPV cell (Supplementary Fig. 14f). The Ta LaBS emitter increases the TPV power density nearly twofold, approaching that of the CFC emitter - but using a material with 3 orders of magnitude lower vapor pressure[18], enhancing TPV lifetime due to lower rates of deposition on the cell[19,66]. Beyond extending lifetime, this increase in power density for a low-vapor pressure material is also critical for the technoeconomics of TPV systems as it reduces the total cell area required[1,14].

Note that in practice the TPV efficiencies and power densities at low temperatures (< 1800°C) could be significantly further improved by using lower bandgap cells designed for those temperatures (as demonstrated in Roy-Layinde et al.[67] and Tervo et al.[65]). Similar relative improvements in power density from using LaBS emitters would then be expected for those devices. This increase in power density is critical for the technoeconomics of TPV systems as it reduces the total cell area required[1,14], potentially halving the cost of electricity produced in certain cost scenarios (e.g. from 30 to 18 ¢/kWh in a system where TPV costs dominate[13,68]). LaBS therefore have the potential to nearly double the profitability of TPV technologies in ideal conditions. In summary, the emissivity measurements (**Fig. 2**) and spectral thermal emission measurements (**Fig. 3**) confirm that LaBS achieve a significant improvement in broadband thermal emission, leading to higher photon emission flux above 1.2 eV and therefore an increased TPV power density.

Increasing TPV power density is only helpful if doing so does not simultaneously reduce its heat-to-electricity conversion efficiency. TPV efficiency, $\eta$, can be calculated by comparing the electrical power generated to the heat absorbed:

$$\eta = \frac{P_{elec}}{P_{elec} + Q_{abs}} \quad (1)$$

where $P_{elec}$ is the electrical power and $Q_{abs}$ is the heat absorbed by the TPV cell, both at the maximum power point. $Q_{abs}$ is measured simultaneously with power using a custom-built heat flux sensor. Detailed experimental calibrations are presented in the Methods section and Supplementary Fig. 18-19. As shown in **Fig. 4h** and Supplementary Fig. 14g, the TPV efficiency using the CFC emitter was 29.07 ± 1.16%, Ta LaBS was 31.79 ± 1.27%, and plain Ta was 31.45 ± 1.44% at 2200°C. There is thus no statistically significant difference in the TPV efficiency between using the Ta LaBS versus the plain Ta emitter, both of which yield comparable TPV efficiencies to using a CFC emitter. This result points to the importance of the TPV cell's Au back-reflector mirror, which significantly reduces parasitic heating from sub-bandgap photons even for the broadband high-emissivity Ta LaBS and CFC emitters. We note that high TPV cell efficiency does not necessarily translate to high system efficiency if this reflected light is unable to make it back to the emitter, such as for open-cavity systems. Therefore, spectrally selective emitters may be necessary for designs lacking good sub-bandgap photon recycling, and related studies have investigated how to use the laser ablation technique presented in this work to design selective emitters[50].

We can use these tests as an indication of high-temperature stability. During the experiments we held each temperature for 10 minutes, and used another 5 minutes to tune the power supply to reach the next temperature, for a total of 90 minutes of high-temperature stability testing from 1700 to 2200°C. We find that compared to the plain Ta emitter, the Ta LaBS exhibits significantly reduced in-band performance degradation after a cumulative 90 minutes of heating at these temperatures, as shown in **Fig. 4i** and Supplementary Fig. 20. The Ta LaBS in-band (above-bandgap) emissivity decreases from 0.982 to 0.947, while the out-of-band (sub-bandgap) emissivity decreases from 0.974 to 0.807 (spectrally weighted at 2200°C).

**Thermal stability of LaBS**

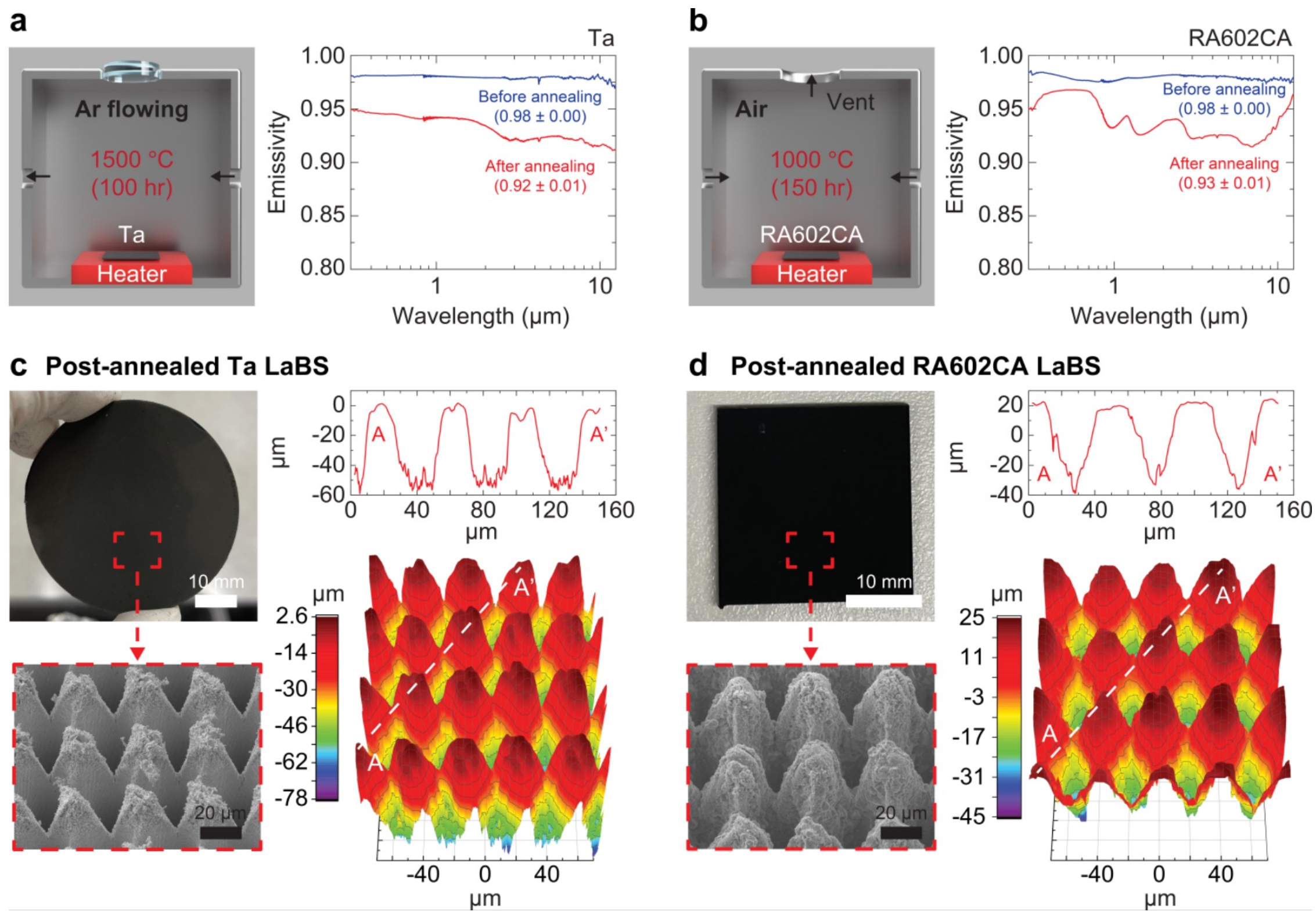

**Figure 5. Thermal stability characterization of Ta and RA602CA LaBS.** (a) Spectral emissivity before and after annealing a Ta LaBS at 1500°C for 100 hours in argon. (b) Spectral emissivity before and after annealing an RA602CA LaBS at 1000°C for 150 hours in ambient air. Error bars represent the range of spectral emissivity across all wavelengths. Surface morphologies characterized by SEM and WLI for post-annealed (c) Ta and (d) RA602CA LaBS. White scale bars are 10 mm, and black scale bars are 20 μm.

Finally, we perform targeted experiments and theoretical modeling to investigate the mechanisms of thermal degradation in LaBS in refractory materials. Thermal stability is critical for practical applications to reduce the number of replacements necessary over a system's lifetime. Thermal stability testing of Ta LaBS (Supplementary Video 6) and RA602CA LaBS are shown in **Figure 5**. These materials are selected because refractory metals are commonly used in TPV systems due to their high melting point above 2500°C, while oxidation-resistant Ni superalloys are often employed in applications where the system operates in ambient air (e.g., CSP)[69].

We subject the Ta LaBS to heating at 1500°C for 100 hours in an argon flowing chamber (**Fig. 5a**, and Supplementary Fig. 21) while the RA602CA LaBS is annealed at 1000°C for 150 hours under ambient air (**Fig. 5b**). After annealing, the Ta LaBS has an average emissivity of 0.92 (between 0.3 μm and 15 μm wavelengths), while RA602CA LaBS has an average emissivity of 0.93, constituting an average emissivity drop of 0.06 or less for both LaBS. SEM images and WLI measurements, as presented in **Fig. 5c-d**, indicate that mesoscale geometries are preserved without clear evidence of thermal damage or deformation on the microstructures. Additionally, the directional spectral absorptivity at different angles up to 1.1 μm wavelength remains above 0.96 after annealing (Supplementary Fig. 22), implying that the ability for the near-omnidirectional light absorption is maintained.

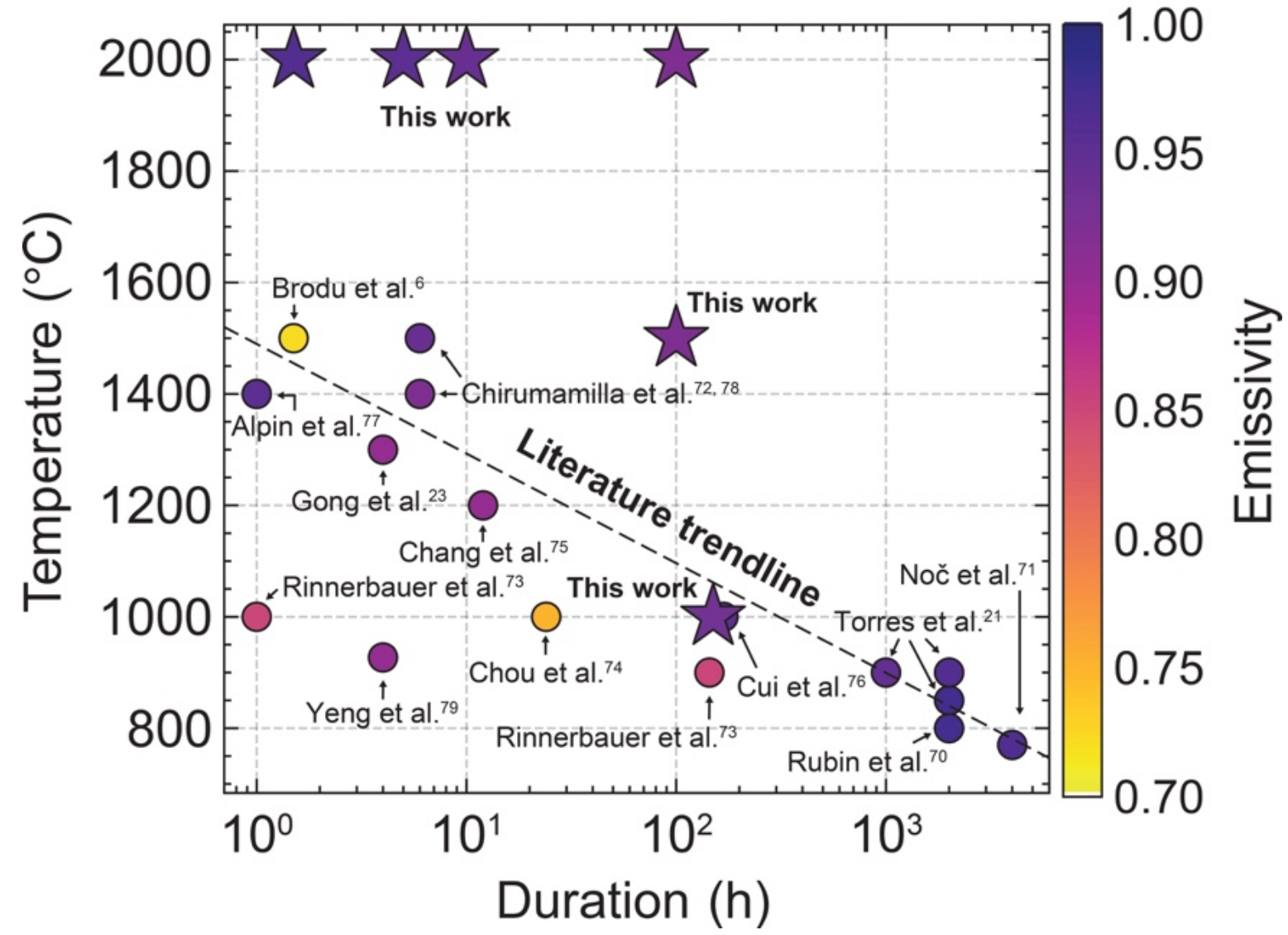

**Figure 6. Comparison with high temperature stability from literature.** Results from stability tests of current state-of-the-art thermal emitters in the literature (circles)[6,21,23,70-79] plotted alongside this work (stars). Each point shows the duration of time that an emitter was held at a specific temperature, and the fill color shows its emissivity at 1 μm measured after the test as indicated by the colorbar. Emissivity at 1 μm is chosen for consistent comparison among the various emitters that included both spectrally selective and broadband emitters. The trendline, fitted only to the state-of-the-art literature data, is a line on these log-linear axes because degradation rates increase exponentially with temperature.

Next, we conducted higher-temperature stability tests using the Ta LaBS serpentine emitters from the previous section, as detailed in Supplementary Fig. 23. We subject the Ta LaBS emitters to heating at 2000°C for 100 hours, in an inert Ar atmosphere with $10^{-19}$ atm $O_2$ partial pressure. Emissivity measurements were taken at 1.5, 5, 10, and 100 hours, as presented in Supplementary Fig. 23b. The in-band emissivity drops to 0.96 after 1.5 hours, 0.95 after 5 hours, 0.94 after 10 hours, and 0.92 after 100 hours. To further demonstrate practicality, we have conducted thermal cycling tests between room temperature and 2000°C, holding at each for 2 minutes with a ramp time of 30 seconds. As shown in Supplementary Fig. 24, the emissivity of the emitter remains high, with in-band emissivity >0.94.

We have plotted our stability results alongside other relevant results from the literature in **Figure 6**. As seen by the literature data fitted trendline, most durability studies are done either at high temperature (> 1200°C) for short durations (< 10 hours), or low temperature (< 900°C) for long durations (> 1000 hours). There is a noticeable gap of emitters (either selective or broadband) that have been characterized for both long durations and high temperatures. Our 100-hour tests at 1500°C and 2000°C meet this gap, and are the highest-temperature durability tests of a structured emitter, to our knowledge.

We believe the micro-/nano- structures formed by fs laser can be robust and durable because they are generated from the substrate itself. In other words, their bulk characteristics (e.g., melting point and oxidation resistance) should be maintained after laser processing, resulting in high thermal stability along with stable long-term radiative properties, if using high-temperature materials. However, we have observed that the micro-/nano- particles attached on Ta microstructures undergo sintering during prolonged heating at elevated temperatures. This sintering causes surface smoothening and decreased particle packing density, as evident from SEM images (Supplementary Fig. 25), leading to a reduction in spectral absorptivity, as illustrated in **Fig. 2g**. As a result, we anticipate that the optical properties of flat Ta with surface roughness will eventually revert to those of the original flat Ta without surface roughness (i.e., low emissivity, as shown in Supplementary Fig. 10b). However, wavy surface structuring provides a second layer of optical property control with larger length scale features that are slower to degrade. Even for locally smooth surfaces lacking surface roughness, high spectral emissivity near a 1 μm wavelength is still preserved for Ta that has these larger length scale wavy microstructures (**Fig. 2g**). Therefore, introducing these wavy microstructures, in addition to the surface roughness provided by the micro-/nano- particles, helps extend the emitter's useful operational lifetime.

To quantify this sintering process, we developed a model to predict how the emissivity would change for long durations or high temperatures (Supplementary Note 3). In a slightly oxidizing atmosphere of $10^{-6}$ atm $pO_2$ (Supplementary Fig. 26a-c), we predict that the Ta LaBS could retain its high broadband emissivity (> 0.96) for up to 100,000 hours at temperatures below 1300°C. At temperatures above 1500°C, shorter lifetimes are anticipated due to surface smoothening caused by the high vapor pressure of $Ta_2O_5$ and its melting point of 1872°C. In contrast, under a reducing atmosphere ($10^{-19}$ atm $pO_2$ as in the TPV measurements, Supplementary Fig. 26d-f), the initial emissivity is expected to be lower because of the early removal of rough $Ta_2O_5$ nanoparticles. Nevertheless, the lifetime at high temperatures is projected to be longer, lasting 10,000 hours at 2000°C, due to the lower vapor pressure and higher melting point of Ta compared to $Ta_2O_5$. The impact of improving durability on cost is presented in Supplementary Fig. 4c.

All in all, these results are promising for applications in TPV and CSP, where traditional surface coatings often fail due to detrimental delamination[69]. By engineering the light absorbing media directly on bulk substrates, we can avoid coating adhesion and delamination issues and achieve near-black emissivity at high temperatures for long lifetimes.

## Conclusions

We demonstrate LaBS with a spectral emissivity exceeding 0.96 across a wide wavelength range across different substrates using ultrafast fs laser ablation, well-suited for enhancing thermal radiative energy transport in energy applications. By decoupling surface optical properties from bulk characteristics, LaBS exhibit thermal stability and enhanced thermal emission, maintaining superior spectral absorptivity after exposure to elevated temperatures exceeding 1000°C for extended periods including in air. Furthermore, TPV power density can more than double when using a LaBS compared to a plain emitter without any loss in efficiency. While the current study focuses on TPV and CSP applications, our scalable approach utilizing ultrafast laser-matter interactions can be easily generalized to various energy harvesting and thermal management applications, including solar-water desalination, passive radiative cooling, and spacecraft, where enhancing radiative energy transport is of particular importance for achieving better performance and efficiency.

## Methods

### Materials

Aluminum, Inconel 600, titanium, stainless steel 301, nickel, graphite, hafnium, and zinc substrates were purchased from GoodfellowUSA. Tantalum substrates were obtained from Sigma Aldrich and GoodfellowUSA and ship with a native oxide layer as confirmed in Supplementary Fig. 27. RA602CA substrates were purchased from Rolled Alloys. Niobium substrates were purchased from Thermo Scientific. Tungsten carbide and molybdenum carbide substrates were available from Stanford Advanced Materials. Detailed specifications of materials are summarized in Supplementary Table 1. As-received materials were used for laser processing without further surface polishing.

### Ultrafast fs laser processing

A 500-fs laser with 1030 nm wavelength operating at 100 kHz repetition rates (s-Pulse, Amplitude) was synchronized with a galvano scanner (excelliSCAN 14, SCANLAB), and *XYZ* stages (A-311 *XY* air-bearing stages with L-310 vertical stage, PI-USA). The focused beam spot size was 30 μm. The processing rate to fabricate LaBS, as shown in **Fig. 1b** and Supplementary Fig. 1b, is approximately 0.072 $cm^2$ per minute.

After fs laser processing, the fabricated samples were sonicated in deionized water for 1 hour to remove weakly adsorbed particles on the surface.

**Optical property characterization**

For visible wavelength ranges (< 0.8 μm wavelength), a UV – VIS spectrophotometer (Lambda 950, Perkin-Elmer) with a 150 mm integrating sphere was used to measure the hemispherical spectral reflectivity. For IR spectral ranges (> 0.8 μm wavelength), a Fourier Transform Infrared spectrometer (Nicolet iS50, ThermoFisher Scientific) equipped with an integrating sphere (Pike technologies) was used. Because none of the substrates are transparent in the IR regime, the spectral absorptivity/emissivity was calculated by '1-reflectivity'. Detailed optical property measurement procedures are provided in Supplementary Note 4 and Supplementary Fig. 6. An accessory component (VeeMax III, Pike technologies) was used to measure the directional specular emissivity.

**Surface morphology characterization**

Scanning electron microscopy (JEOL), white light interferometry (NewView 6000, Zygo), and laser confocal microscopy (VK-X3000, Keyence) were used to examine the surface morphology.

**FDTD simulation**

FDTD simulations were performed using Lumerical Inc. software. The simulated structure, shown in **Fig. 2f**, features domain sizes of 20 μm in both the *X* and *Y* directions, with an amplitude of 40 μm and a thickness of 5 μm. The mesh size is set to 50 nm in the *X* and *Y* directions and 100 nm in the *Z* direction. Periodic boundary conditions are applied in the *XY* plane, while perfectly-matched-layer boundary conditions are used in the *Z* direction. Different surface roughnesses are generated using the built-in script (ID: rough_surf), maintaining a consistent correlation length of 0.1 μm, but varying the root-mean-squared amplitudes. The auto-off value is set to $1\times10^{-6}$. The simulated spectral emissivity is derived from the simulated spectral reflectivity (i.e., emissivity = 1 - reflectivity). We conducted the FDTD simulations using a workstation equipped with an Intel Xeon CPU E5-2687W v3 (3.10 GHz) and 64 GB of RAM. Under these conditions, each simulation was completed in approximately 12 hours.

**Thermal emission measurement**

Thermal emission was measured by an IR camera (Spark M150, Telops) and spectral bandpass filters (Iridian Spectral Technologies) with central wavelengths at 1.9 μm, 2.7 μm, 3.46 μm, 4.26 μm, and 4.46 μm. Samples were heated by a heater (Model #101491, HeatWave Labs) in an argon gas flowing chamber (Praxair, 99.999% ultrahigh purity), and the temperature was measured by an embedded K-type thermocouple.

**TPV measurement**

The TPV cells used are a 1.4/1.2 eV tandem GaAs/GaInAs cell with a Ag back surface reflector (MT618, National Renewable Energy Laboratory) and a 1.2/1.0 eV tandem AlGaInAs/GaInAs cell with a Au back surface reflector (MT671, National Renewable Energy Laboratory). Detailed schematics of the TPV cell structures are presented in LaPotin at el.[1]. We use III-V tandem cells with a back-reflector to improve performance by reducing thermalization and parasitic heat absorption. Thermophotovoltaic power is measured by a 4-wire measurement (Keithley) using electrical leads placed on the top and bottom contacts of the cell. At each temperature, an IV curve is obtained using the 4-wire measurement, and the maximum power point is determined. At that maximum power point, the electricity produced and heat absorbed are measured and efficiency is calculated.

Heat absorbed is measured with a custom-built heat flux sensor that is calibrated according to prior methodology[64]. The custom heat flux sensor consists of a copper bar with 4 thermocouples embedded in it. The bottom of the copper bar is kept at 5°C using a thermoelectric, and the TPV cell is attached to the top with thermal paste.

From the electrical power output and heat absorption values, the efficiency can be calculated (using Equation 1), but there are sources of superfluous heating from the measurement apparatus that must be accounted for. Although the electrical lead placement is designed to minimize superfluous heating by conduction to the emitter, some heat absorption by the measurement leads is inevitable, which would cause an underestimate of efficiency[80]. To correct for this superfluous absorption, at each considered temperature the heat absorbed is measured with ($Q_{voc,leads}$) and without ($Q_{Voc,no\ leads}$) the electrical leads in place, both at open circuit conditions. The difference in heat absorbed is thus the contribution of superfluous heating of the leads and is plotted in Supplementary Fig. 18. Another source of superfluous heating is Joule heating at the contact of the electrical lead to the cell[80], so the contact resistance ($R_{contact}$) was measured to be 0.092 ohms and the extra heating was accounted for with $I^2 R_{contact}$, where $I$ is the current generated by the TPV cell at its maximum power point. Therefore, the calibrated heat absorption is calculated by subtracting these superfluous sources of heat from the total heat measured by the heat flux sensor ($Q_{HFS}$):

$$Q_{abs} = Q_{HFS} - (Q_{Voc,leads} - Q_{Voc,no\ leads}) - I^2 R_{contact} \tag{2}$$

After conducting this calibration and calculating the efficiency, we note that the values are significantly less than those reported in LaPotin who reported a 37% efficiency using the same experimental setup with a CFC emitter[64]. Therefore, it is likely the cells used in this work are lower quality with worse quantum efficiency and higher series resistance than the best quality cells of the same architecture. Future experiments could use higher quality cells and achieve better efficiency.

The measurement was conducted in a chamber with oxygen partial pressure controlled to $10^{-18}$ atm to limit the oxidation of the Ta sample. To achieve these low oxygen partial pressures, a crucible of Zr powder (Strem) was heated to > 600°C inside the chamber, reacting with the oxygen to form zirconia.

The emitter was heated with Joule heating by supplying a current with a power supply (MagnaDC). To match the electrical resistance properties of the emitter with the specifications of the power supply, the CFC emitters were made as a monolithic piece while the Ta emitters had a serpentine pattern, as shown in Supplementary Fig. 15 and Supplementary Video 5. However, the monolithic piece had the same view factor as the serpentine path, as verified in both simulation and experiment (Supplementary Fig. 19). The emitters were placed 5 mm above the TPV cell to achieve a high view factor of 0.39, with an aperture plate placed in between to limit parasitic heat absorption in components around the cell.

This view factor most closely corresponds to a system configuration where the emitter area is smaller than the cell area (e.g. concentric cylinders with cells on the outer cylinder). These configurations are typically characterized by low power density, but we have shown the LaBS emitters can help offset this limitation and achieve higher power density due to their higher emissivity. Other studies have examined higher view factors in configurations where the emitter and cells are closer together or where the emitter fully surrounds the cells. These studies found that while higher view factors tend to reduce efficiency (due to increased series losses), they increase power density by increasing the current density[81]. We note that increasing emitter emissivity would enhance power density for any view factor configuration, if holding all other parameters constant.

The temperature of the emitter was measured using a two-color pyrometer (Fluke Endurance E1RH) placed above the emitter. The top surface of the Ta emitters was LaBS to ensure constant emissivity in the two measurement bands (centered at 0.95 and 1.05 μm wavelength). The emitter was held at each temperature for 10 minutes to ensure a thermal steady state was reached. IV curves were taken after 5 minutes at each temperature, and the power and heat absorption values were averaged over the last 5 minutes of each temperature.

**Thermal stability test**

A box furnace (Lindberg/Blue M 1100°C, Thermo Fisher Scientific) was used to anneal RA602CA substrates in ambient air. A custom sublimation chamber built by Antora Energy Inc. was used to anneal Ta substrates at 1500°C under argon gas condition.

**Acknowledgement**
Authors thank Antora Energy Inc. for use of one of their sublimation chambers. The authors thank Myles Steiner and the Chemistry and Nanoscience Center at the National Renewable Energy Laboratory for providing the TPV cells used in the study. The authors thank Duo Xu and the Boriskina Research Lab / NanoEngineering Group for lending the integrating sphere used for part of the reflectance measurements in this work. This work was carried out in part through the use of MIT.nano's Characterization facilities.

**Funding:** This work was supported by the Laboratory Directed Research and Development program of Lawrence Berkeley National Laboratory under U.S. Department of Energy Contract No. DE-AC02-05CH11231, and the ARPA-E Contract No. 2107-1539 to Lawrence Berkeley National Laboratory. This work was partially supported by Solar Energy Technologies Office (SETO) under U.S. Department of Energy Contract No. DE-EE0009819. This work was supported by the National Science Foundation Graduate Research Fellowship under Award No. 2141064.

**Author contributions:**
Conceptualization: M.P., S.V., A.L., R.P., A.H., S.D.L., C.P.G., and V.Z.
Methodology: M.P., S.V., A.L., D.P.N., R.P., A.H., S.D.L., C.P.G., and V.Z.
Investigation: M.P., S.V., and A.L.
Supervision: R.P., A.H., S.D.L., C.P.G., and V.Z.
Writing-original draft: M.P., S.V., A.L., R.P., A.H., S.D.L., C.P.G., and V.Z.
Writing-review and editing: M.P., S.V., A.L., R.P., A.H., S.D.L., C.P.G., and V.Z.

**Conflict of interest**
The authors declare no competing interests.

**Data availability**
The authors declare that the data supporting the findings of this study are available within the article and its Supplementary Information.